\documentstyle[mprocl]{article}

\input{psfig}
\def\Journal#1#2#3#4{{#1} {\bf #2}, #3 (#4)}

\def\PRL{\em Phys. Rev. Lett.}
\def\PRD{{\em Phys. Rev.} D}

\def\PR{\em Proc. R. Soc. London}

\def\be{\begin{equation}}
\def\ee{\end{equation}}
\def\bea{\begin{eqnarray}}
\def\eea{\end{eqnarray}}

\begin{document}

\title{STRUCTURE OF THE CAUCHY HORIZON SINGULARITY}

\author{LIOR M. BURKO}

\address{Department of Physics, Technion---Israel Institute of
Technology, 32000 Haifa, Israel}




\maketitle\abstracts{
We study the Cauchy horizon (CH) singularity of a
spherical charged black hole perturbed nonlinearly by a
self-gravitating massless scalar field. We show numerically
that the singularity is weak both at the early and at the late
sections of the CH,
where the focusing of the area coordinate $r$ is strong.
In the early section the metric perturbations vanish, 
and the fields behave according to perturbation analysis.
We find exact analytical expressions for the gradients
of $r$ and of the scalar field, which are valid at both
sections. We then verify these analytical results numerically.}

\section{Introduction}
In the last few years, there has been a steadily growing evidence
that a new type of singularity forms at the Cauchy horizon (CH)   
of spinning or charged black holes (BHs). The features of this new
singularity
differ drastically from those of the previously-known
singularities like, e.g., Schwarzschild or BKL: First, the CH
singularity is null rather than spacelike; Second,
it is weak \cite{ori1}. Namely, the tidal distortion experienced by
an infalling extended test body is finite (and, moreover,
is typically negligibly small) as it hits
the singularity. Yet, curvature scalars diverge there
\cite{pi}. 

Despite recent advances \cite{ori2,ori-grg,flanagan,bc,israel}, 
our understanding of the null weak CH
singularity is still far from being complete.
In particular, it is important to verify this new picture by
performing independent, non-perturbative, analyses. This motivates
one to employ numerical tools to
study the structure of the CH singularity. The numerical simulation of
spinning BHs is difficult, as they are non-spherical. One is thus led
to study, numerically, the inner structure of a spherical charged
BH; hopefully, it may serve as a useful toy model for a spinning BH.
 
Recently, Brady and Smith (BS) \cite{brady} numerically explored
the mass-inflation singularity inside a spherical charged BH perturbed by
a spherical scalar field. This analysis
confirmed several aspects of the above new picture:
It demonstrated the existence of a null singularity
at the CH, where the mass function $m$ diverges but the area coordinate   
$r$ is nonzero.
$r$ was found to decrease monotonically, due to the nonlinear focusing,
until it shrinks to zero (at which point the singularity becomes
spacelike). It also 
provided evidence for the weakness of the singularity.
Despite its remarkable achievements, however,
this analysis left one important issue unresolved:
To what extent is the
perturbative approach applicable at (and near) the CH singularity?
BS reported on an inconsistency with the predictions of perturbation
analysis,
manifested by the non-zero value of $\sigma$ (see \cite{brady}).
This issue is crucial, because for realistic (i.e., spinning and 
uncharged) BHs
the only direct evidence at present for the actual occurrence of a null
weak
singularity stems from the perturbative analysis \cite{ori2}. A failure of  
the perturbative approach in the spherical charged case would therefore
undermine the confidence in our understanding of realistic BHs' interiors.
In the next section we briefly
describe our numerical and analytical results. 
Further details on this research will 
appear elsewhere \cite{burko1}.

\section{The numerical and analytical investigations} \label{sec2}
We consider the model of a spherical charged BH 
perturbed non-linearly by a spherical, self-gravitating, neutral,
massless scalar field $\Phi$. Our numerical code is based on free 
evolution of the dynamical equations in double-null coordinates
\cite{burko2,burko3}. The code is stable and second-order accurate. 
Our initial-value setup is described in Ref. \cite{burko2}:
The geometry is initially Reissner-Nordstr\"{o}m (RN),
with initial mass $M_0=1$ and charge $Q$, and no scalar field.
At some retarded moment $v$, however, it is modified by an ingoing
scalar-field pulse of
a squared-sine shape with amplitude $A$. $\Phi$ vanishes everywhere on
the initial surface except in a finite range $v_1<v<v_2$.
In this case, due to the scalar-field energy, the BH's external mass
approaches the final mass $M_f$.
Note that our outgoing initial null hypersurface is located
outside the event horizon (EH) (unlike in Ref. \cite{brady}).

Our numerical simulations confirm the presence of a null
singularity at the CH, where
$m$ diverges and $r$ is nonzero. Along the CH singularity, $r$
decreases monotonically, until it shrinks to zero, at which point the
singularity becomes spacelike.
This situation was already found numerically by BS \cite{brady}.
We next investigate the early part of the CH singularity, i.e.,
the part where the focusing of
$r$ is still negligible. Our first goal is to demonstrate that 
the singularity is weak.
In terms of the double-null metric, the singularity 
will be weak if
coordinates $\hat u(u),\hat v(v)$ can be chosen such that
both $r$ and $g_{\hat u \hat v}$ are finite and
nonzero at the CH.
The numerical analysis by BS already demonstrated the finiteness of $r$,
which
we recover in our results. (Note that $r$ is independent of the choice of
the null coordinates.) Figure 1(A)
displays the metric function  $g\equiv -2g_{\hat u \hat v}$ in
Kruskal-like
coordinates $U,V$ along an outgoing null ray
that intersects the early section of the CH singularity.
The CH is located at $V=0$ (corresponding
to $v\to\infty$). This figure clearly demonstrates the finiteness of
$g_{UV}$, from which the weakness of the singularity follows. 
The perturbation analysis also predicts that both
the scalar field and the metric perturbations will be
arbitrarily small at the early section of the
CH. In other words, both metric functions $r$ and $g$
should be arbitrarily close to
the corresponding RN metric functions. 
This behavior is indeed demonstrated in Fig. 1.
In this figure, $g$ and $r$ are displayed along lines $u={\rm const}$
[Figs. 1(A) and 1(B)]
and $v={\rm const}$ [Figs. 1(C) and 1(D)]. 
The RN values are denoted by solid lines, and the numerical values in
circles. 
The similarity of the analytic RN functions and the
numerically-obtained functions of the perturbed spacetime is remarkable.
(We emphasize, though, that despite the similarity in the values of the
metric functions to RN, our geometry is drastically 
different from that of RN, in the sense
that in our case curvature blows up at the CH, as manifested by the
rapid growth of $m$ [Fig. 1(E)].)
 
We turn now to explore the late part of the CH singularity, where
focusing is strong.
First, we numerically verify the weakness of the singularity in this part
too.  
Figure 1(F) shows $K\equiv -2g_{uV}$ along an outgoing null ray in the
late
part of the CH, where the focusing level (of $r$) is approximately $90\%$.
We present here the results for various values of the grid-parameter
$N$ (see \cite{burko2}), in order to demonstrate the second-order
numerical convergence.
$g_{uV}$ approaches a finite value at the CH ($v\to \infty$).
At the same time, 
the mass function (and curvature) grows exponentially with $v$
[Fig. 1(G)]. We therefore conclude
that the entire null CH singularity is weak, 
even at the region of strong focusing.

Next we study, analytically, the behavior of the blue-shift factors
$r_{,v}$ and $\Phi_{,v}$
along the contracting CH. The field equations can 
be integrated exactly along the CH singularity. We find
$R\equiv \Psi_{,v}/\{\Psi_{,v}^{\rm EH}[2(M_f/Q)^{2}-1]\}=1$ and 
$P\equiv -{\left(r^{2}\right)}_{,v_{e}}/\left[
(1/\kappa_{-}){\Psi_{,v_{e}}}^{2}\right]=1$.
Here, $\kappa_{-}$ is the surface gravity at the RN inner horizon with 
parameters $M_f$ and $Q$, and $\Psi\equiv r\Phi$.
Note that $P$ is invariant to a 
gauge transformation $v\to \tilde{v}(v)$,
whereas $R$ is not.
In order to verify these expressions, we calculated numerically and
plotted 
$R$ (dashed) and $P$ (solid), as functions of $v$,
along an outgoing ray located at a region of $90\%$ focusing of the CH.
The results, presented in Fig. 1(H), are in
excellent agreement with the above theoretical prediction, $R=1$
and $P=1$.
\begin{figure}
\rule{6cm}{0.2mm}\hfill\rule{6cm}{0.2mm}
\psfig{figure=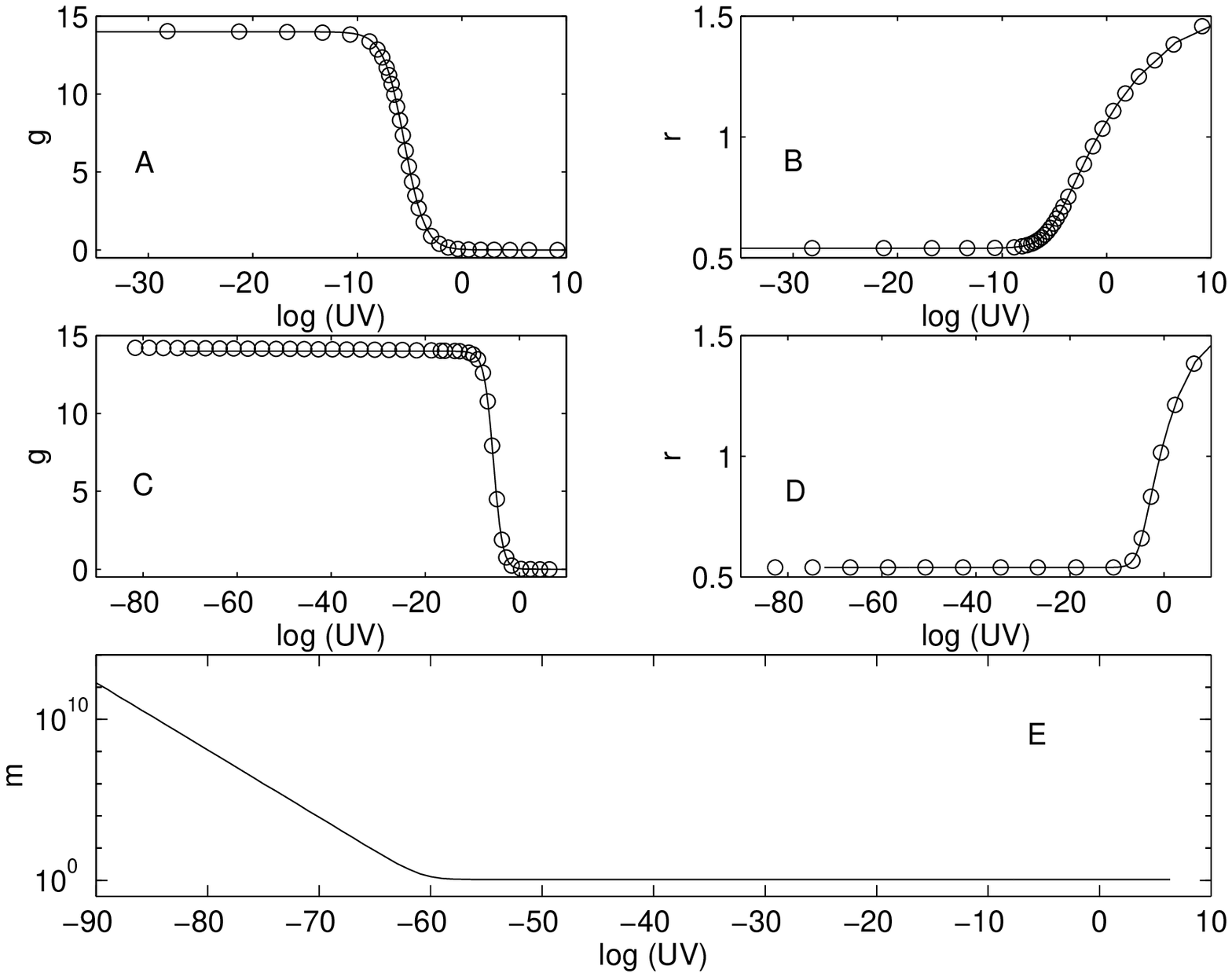,height=2.0in}
\hfill
\psfig{figure=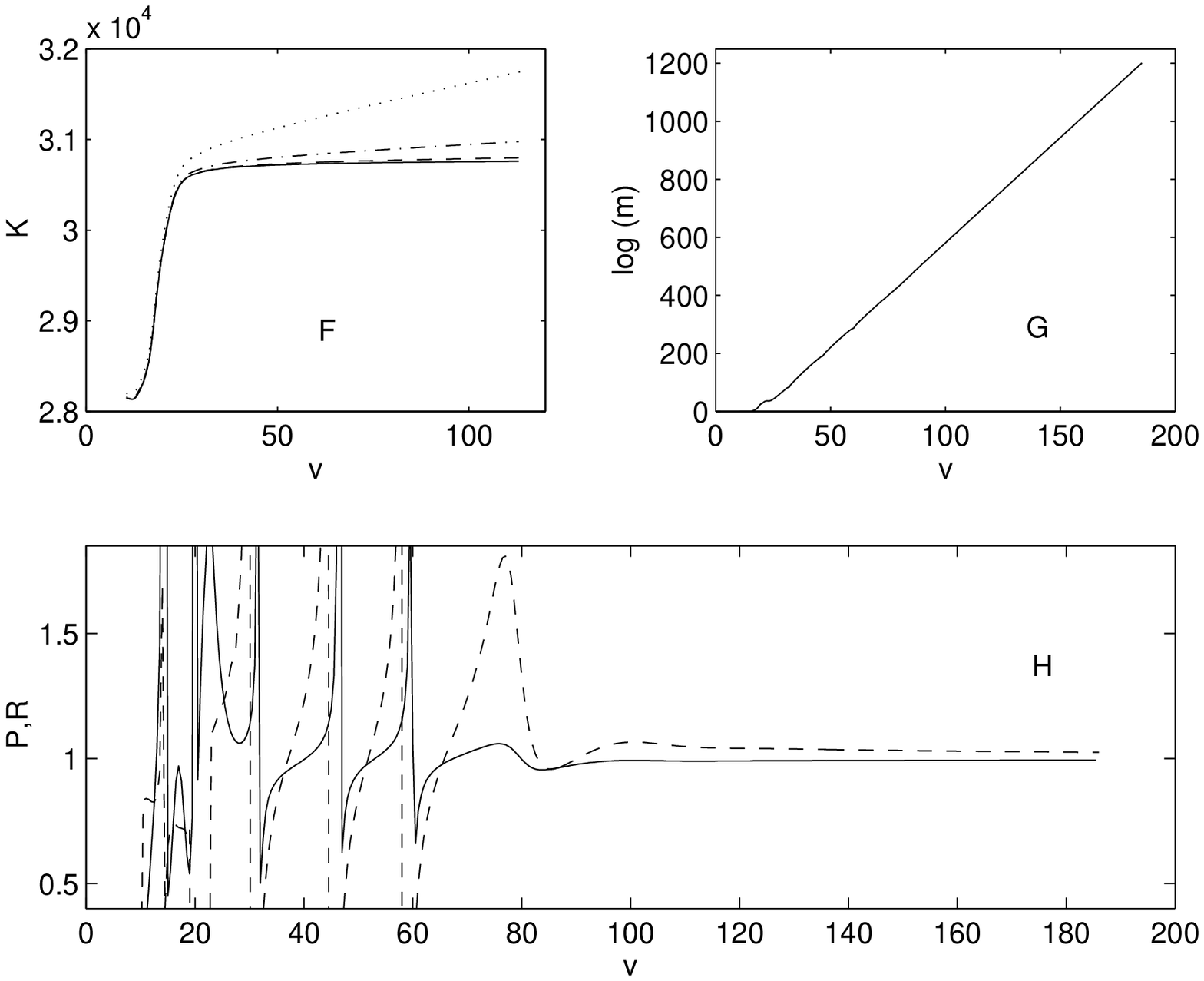,height=2.0in}
\caption{Right: The metric functions and the mass function at the early
parts of the CH. Left: metric, mass function and normalized blue-shift
factors at the late parts of the CH.
\label{fig:radish}}
\end{figure}
\section*{Acknowledgments}
I thank Amos Ori for many stimulating and helpful discussions.

\end{document}